\documentstyle[preprint,aps]{revtex}
\begin{document}
\draft
\def\ket#1{| #1 \rangle}
\def\bra#1{\langle #1 |}
\def\R{{\cal R}}
\def\Ha{{\cal H}_A}
\def\He{{\cal H}_{\rm e}}
\def\Hc{{\cal H}_{\rm c}}
\def\Hq{{\cal H}_{\rm q}}
\def\tto{\rightarrow}

\title{Wave-function entropy and dynamical-symmetry \\
breaking in the interacting boson model}
\date{\today}
\author{Pavel Cejnar$^1$ and Jan Jolie$^2$}
\address{$^1$Dept. of Nuclear Physics, Charles University,
V~Hole\v sovi\v ck\'ach 2, \\ CZ---180\,00 Prague, Czech Republic\\
$^2$Dept. of Physics, University, P\'erolles, CH---1700 Fribourg,
Switzerland}
\maketitle

\begin{abstract}
The degree of chaos in the Interacting Boson Model (IBM-1) is compared
with what we call the \lq\lq dynamical-symmetry content\rq\rq\ of
the system.
The latter is represented by the information entropy of the
eigenfunctions with respect to bases associated with dynamical
symmetries of the IBM-1, and expresses thus the localization of actual
eigenfunctions in these symmetry bases.
The wave-function entropy is shown to be a sensitive tool for monitoring
the processes of a single dynamical-symmetry breaking or transitions
between two and more symmetries.
For the IBM-1 hamiltonians studied here, the known features related to
chaos, namely the dependence of chaotic measures on the hamiltonian
parameters (position in the Casten triangle) and on the angular momentum,
turn out to be correlated with the behaviour of the wave-function
entropy.

\pacs{21.60.Fw; 05.45.+b}
\end{abstract}

\section{Introduction}
\label{introduction}

Although fundamental quantum-mechanical equations of motion are
about seventy years old, the task to understand the whole variety
of phenomena \lq\lq encoded\rq\rq\ in them has not been completed
yet.
One of the most interesting problems of this sort is to find quantum
signatures of the order-to-chaos transition:
If the dynamics of a classical system is changed, by varying some
parameters, from the regular to chaotic regime, what happens on the
level of the system's quantum counterpart\,?
Much insight into the classical--quantum correspondence has been
gained in connection with this question in recent years
\cite{Gutzwiller}, but many problems still remain open.

It is well known that the order/chaos signatures can be found in
both the factors that determine quantum dynamics of any bound system,
i.e., in both discrete sets of (i) energy eigenvalues and (ii)
corresponding eigenfunctions.
The transition to chaos in the classical limit was found to set
up specific short- and long-range correlations in the energy spectrum
and a regime of Gaussian overlaps of the energy eigenfunctions with
any probe state \cite{Guhr}.
Besides these properties, inherently described by random--matrix
theory \cite{Mehta}, also certain spatial and temporal properties of
wave functions (such as nodal-line patterns, scars, wave-packet
dynamics etc.) were shown to be affected by the order-to-chaos
transition \cite{Gutzwiller}.

Regularity and chaos are sometimes thought as limiting manifestations
of various degrees of symmetry contained in the system
\cite{Feshbach}.
Indeed, regularity is obviously related to integrability (integrable
systems are always totally regular) and the latter, since it
ensures a number of compatible integrals of motion, is nothing else
than \lq\lq a kind of symmetry\rq\rq.
When dealing with symmetries, we do not have in mind exact symmetries
that the system can exhibit, but rather the generalized, so-called
dynamical symmetries \cite{Castanos,Frank,Rowe,Zhang1,Zhang2}.
Although dynamical symmetries are defined only for algebraic systems
(i.e., systems associated with a dynamical group G whose
representations define the system's Hilbert space and whose generator
algebra induces all the relevant operators on this space), their role
in physics is probably quite general.

A system with a dynamical group G has a dynamical symmetry if its
hamiltonian can be written exclusively in terms of Casimir operators
of the subgroups involved in the chain
\begin{equation}
{\rm G}\supset {\rm G}_1 \supset \dots \supset
{\rm G}_i \supset \dots \supset {\rm G}_k\ ,
\label{dynsym}
\end{equation}
which specifies the given dynamical symmetry.
Note that only the smallest embedded group, G$_k$, is an ordinary
symmetry group.
The systems with dynamical symmetries have two rather exclusive
properties:
First, their eigenspectra and corresponding eigenfunctions can be
found analytically \cite{Frank} and, second, they are integrable
\cite{Zhang1,Zhang2,Zhang3,Zhang4}.
However, also known is that, in turn, integrability {\it does not\/}
have to always couple with dynamical symmetries, which means that
the link of the present concept of symmetry to chaos is still
imperfect.
 
In spite of the above-mentioned numerous signatures of quantum
chaos found in recent years, very little has been said about
the connection of the quantum order-to-chaos transition with
the process of dynamical-symmetry breaking.
In particular, the question should be addressed \lq\lq to what
extent\rq\rq\ a concrete dynamical symmetry must be broken to
induce the transition to chaos \cite{Zhang4}.
An obvious problem is that whereas dynamical symmetry (and
integrability) follows a simple boolean logic\,---\,the system
either has it or not\,---\,the order-to-chaos transition is
a rather smooth affair:
Perturbations of an integrable hamiltonian usually do not
immediately bring the system to a completely chaotic regime,
but make it pass some transitional region.
The explanation of this behaviour in the classical case was the
main goal of the famous KAM theorem \cite{Gutzwiller}.
However, what {\em in the dynamical-symmetry language\/} controls
the degree of quantum chaos\,?

In this paper we would like to show that not only the degree of
chaos inherent in the system, but also its \lq\lq dynamical-symmetry
content\rq\rq\ can be measured by a continuous quantity.
Our approach is based on the fact that any particular dynamical
symmetry is associated with a certain basis (or a subclass of bases
in general).
This basis, obtained by the simultaneous diagonalization of
Casimir operators of all groups in the dynamical-symmetry chain,
becomes a reference frame in the system's Hilbert space.
The degree of localization of actual eigenstates in the reference
basis, i.e., the degree of overlap of the two bases then
represents the desired continuous measure which tells us
\lq\lq how close\rq\rq\ to the given integrable case the actual
system is.
In this logic, chaos is smoothly established as the degree of
eigenstate localization in the bases corresponding to all
possible dynamical-symmetries of the system decreases.

To illustrate these general ideas, we invoke the simplest version
(IBM-1) of the Interacting Boson Model \cite{Iachello}, which
is known to efficiently describe basic aspects of collective
motion in atomic nuclei.
Apart from the advantage that the IBM group structure is explicitly
known, the model also sets a field for practical applications of our
investigations in realistic systems.
To quantify the eigenstate localization of a particular transitional
IBM-1 hamiltonian in a given dynamical-symmetry basis, we use the
so-called information entropy of wave functions.
It will be shown that the entropy analysis of eigenfunctions
enables  one to find a counterpart of all order/chaos properties of
the IBM-1 in terms of the known dynamical symmetries associated with
this model.

Here is the plan of the paper:
Properties of the particular form of the IBM-1 used are
discussed in Sect.\ref{model}, while general features of the
wave-function entropy are described in Sect.\ref{entropy}.
In Sect.\ref{results}, we present results of the entropy calculations
for various transitional IBM-1 hamiltonians in dependence on the
angular momentum and boson number, and show their relation to the
chaotic properties of the model.
Sect.\ref{conc} brings a few concluding remarks.

\section{Model}
\label{model}

\subsection{Dynamical group}
\label{dyngroup}

As already indicated above, in this work we will study the
wave-function entropy within the Interacting Boson Model--1. 
The IBM-1 was introduced in 1974 by Arima and Iachello with the
aim to describe collective nuclear excitations in an algebraic
framework (see references in Refs.\cite{Frank,Iachello}) and has
soon since evolved into more sophisticated and powerful boson
models \cite{Frank}.
Because we do not make a specific link to nuclear physics here
(although some of the results may turn out applicable there),
we use the original and simplest version of the IBM.

The IBM-1 is formulated as a model describing one- and two-body
interactions of two kinds of bosons, named s- and d-bosons according
to their spins $l$=0 and 2, respectively.
Because the interactions conserve the total number of bosons $N$,
the dynamical group of the model is the U(6) generated by the
36 bilinear products of the boson creation and annihilation
operators:
$s^{\dagger}s$, $d_{m}^{\dagger}d_{m'}$, $d_{m}^{\dagger}s$ and
$s^{\dagger}d_{m}$ ($m,m'=-2\dots+2$).
The hamiltonian, built only from these products, is further made
invariant under time reversal and rotations by allowing for
only the hermitian terms with zero total angular momentum.
Carrier spaces of irreducible representations of the dynamical
group, each of them corresponding to a fixed boson number $N$,
naturally coincide with possible quantum Hilbert spaces ascribed
to the model.
If the boson operators in the IBM-1 hamiltonian are rewritten in
a convenient coordinate representation, the model turns out to
describe rotations and quadrupole vibrations of a specific quantum
\lq\lq drop\rq\rq.

\subsection{Dynamical symmetries and integrability}
\label{symmetries}

Possible dynamical symmetries of the IBM-1 are found by constructing
various subgroup chains of the dynamical group U(6), all ending at the
angular-momentum group SO(3) generated by the products
$[d^{\dagger}\times{\tilde d}]^{(1)}_{\mu}$ (standard definitions
$[b_l^{\dagger}\times{\tilde b}_{l'}]^{(\lambda)}_{\mu}=\sum_{mm'}
(lml'm'|\lambda\mu)b_{lm}^{\dagger}{\tilde b}_{l'm'}$ and
${\tilde b}_{lm}=(-1)^{l+m}b_{l-m}$ are used), which must remain the
symmetry group of the hamiltonian.
Such, one finds the following three chains \cite{Frank,Iachello}:
\begin{eqnarray}
{\rm (I)}\quad & {\rm U(6)} & \supset{\rm U(5)}\supset{\rm SO(5)}
\supset{\rm SO(3)}\ ,\nonumber\\
\label{chains}
{\rm (II)}\quad & {\rm U(6)} & \supset{\rm SU(3)}\supset{\rm SO(3)}\ ,\\
{\rm (III)}\quad & {\rm U(6)} & \supset{\rm SO(6)}\supset{\rm SO(5)}\supset
{\rm SO(3)}\ .\nonumber
\end{eqnarray}
Note that the concrete realization of some groups in (\ref{chains}) is
not unique because of phase ambiguities of the boson operators.
We will come to that point later.

A possible set of Casimir operators ${\hat{\cal C}}_i[{\rm G}]$ of the
first and/or second order ($i$=1 and/or 2, respectively) of the groups
G involved in the chains (\ref{chains}) is listed in Tab.\,I (dots
denote the scalar product defined by $A^{(\lambda)}\cdot B^{(\lambda)}
=\sum_{\mu}(-1)^{\mu}A^{(\lambda)}_{\mu}B^{(\lambda)}_{-\mu}$).
The U(6) invariants were skipped because, as already mentioned above,
within the IBM-1 one always takes into account only one
finite-dimensional subspace of the total Hilbert space corresponding
to a fixed (sharp) boson number $N$, where the U(6) Casimir operators
yield just ordinary numbers.
The most general IBM-1 hamiltonian can be written as a linear
superposition (weighted sum) of the invariants from Tab.\,I and, as
such, it has 6 free parameters\,---\,the weights.
If all the weights are zero except for those standing at invariants of
only one group chain in (\ref{chains}), the hamiltonian has the
dynamical symmetry described by the given chain.
Hereafter, we will distinguish these dynamical symmetries by the name
of the corresponding largest subgroup, i.e., by labels U(5), SU(3)
or SO(6), respectively.
These limits correspond to vibrational, rotational and
$\gamma$-unstable nuclei \cite{Frank,Iachello}.

The above-specified symmetries do not, however, exhaust all possible
dynamical symmetries of the model.
Namely, phase ambiguities in the definition of boson operators lead
to two additional symmetries connected with the following chains
\cite{Frank,Iachello,Isacker,Kusnezov,Cejnar}:
\begin{eqnarray}
{\rm (II*)}\quad & {\rm U(6)} & \supset{\rm SU(3)*}\supset{\rm SO(3)}\ ,
\nonumber\\
{\rm (III*)}\quad & {\rm U(6)} & \supset{\rm SO(6)*}\supset{\rm SO(5)}
\supset{\rm SO(3)}\ .
\label{chains*}
\end{eqnarray}
The group SU(3)* is made from the \lq\lq standard\rq\rq\ SU(3) by the
transition $(d^{\dagger}_m,{\tilde d}_m)\to(-d^{\dagger}_m,-{\tilde
d}_m)$ (which is equivalent to taking $\chi$=$+\sqrt{7}/2$ instead
of $-\sqrt{7}/2$ in ${\hat{\cal C}}_2[{\rm SU(3)}]$, see Tab.\,I) and
SO(6)* is made from SO(6) by $(d^{\dagger}_m,{\tilde d}_m)\to(-id
^{\dagger}_m,i{\tilde d}_m)$ (equivalent to $\phi$=0 instead of $\pi$
in ${\hat{\cal C}}_2[{\rm SO(6)}]$).
The SU(3)* and SO(6)* Casimir operators can be written as linear
superpositions of the Casimir operators in Tab.\,I \cite{Cejnar}.
Therefore, the hamiltonian has dynamical symmetry SU(3)* (II*)
or SO(6)* (III*) if some of the weights in its expansion have certain
ratios.
In particular \cite{Cejnar}, the SU(3)* dynamical symmetry sets in
if ratios of the coefficients at Casimir operators (see Tab.\,I)
${\hat{\cal C}}_1[{\rm U(5)}]$, ${\hat{\cal C}}_2[{\rm U(5)}]$,
${\hat{\cal C}}_2[{\rm SO(6)}]$, ${\hat{\cal C}}_2[{\rm SO(5)}]$,
and ${\hat{\cal C}}_2[{\rm SU(3)}]$ are $2:2:4:-6:-1$, whereas
the SO(6)* dynamical symmetry requires the ratios $4(N+2):-4:-1:
{\rm arbitrary}:0$.

Since all phase conventions must fulfil restrictions following
from rotational and time-reversal invariance of the resulting
hamiltonian, the above five chains most probably represent
a complete set of the IBM-1 exact dynamical symmetries.
This is not so, however, as far as the integrability of the
model is concerned.
The IBM-1 is, in fact, a straightforward example showing that
although a dynamical symmetry really implies integrability,
the opposite implication (proposed in Ref.\,\cite{Zhang2}
but soon abandoned \cite{Zhang4}) does not, in general, hold.
To see that, consider the IBM-1 hamiltonian which is
transitional between the U(5) and SO(6) dynamical symmetries,
but has no admixture of the SU(3) invariant.
Neither the U(5) nor the SO(6) invariant is an integral of
motion on the U(5)--SO(6) transition, but the hamiltonian
itself, which is always a trivial commuting integral of motion,
becomes independent from the other integrals, which ensures
that the integrability is preserved if the SO(5)-generated
dynamical symmetry is taken into account
\cite{Alhassid2,Whelan1}.
This is a special case of the self-evident rule that
non-integrability does not appear if the dynamical-symmetry
breaking destroys just one integral of motion \cite{Zhang4}.
For instance, if apart from the chain (\ref{dynsym}) also
${\rm G}\supset {\rm G}_1 \supset \dots \supset {\rm G'}_i
\supset \dots \supset {\rm G}_k$ (differing only by the $i$-th
subgroup) is a valid group reduction, the transition between
these two dynamical symmetries is always totally integrable,
although having no dynamical symmetry in the above sense.

\subsection{Casten triangle}
\label{castentri}

The six free parameters of the most general IBM-1 hamiltonian is
a too large number if features of the model are to be systematically
scanned over the whole parameter space.
Nevertheless, it is often enough to select a certain subset of
hamiltonians out of the complete set generated by the dynamical
group.
Usually a two-dimensional manifold is spread in the six-dimensional
parameter space so that the U(5), SU(3), and SO(6) limits are reached
for some particular points.
The manifold is then mapped onto the so-called Casten triangle, whose
vertices correspond to these limits.

Chaotic properties of a specific two-parameter set of IBM-1
hamiltonians were throughoutly studied by Alhassid, Whelan, and
Novoselsky \cite{Alhassid2,Whelan1,Alhassid1,Alhassid3,Alhassid4}.
As we follow the cited works in order to relate the order/chaos
signatures to the dynamical-symmetry content, we use the same
parameterization of the IBM-1 hamiltonian.
It is given by the following formula (see Tab.\,I):
\begin{equation}
{\hat H}_{(N,\eta,\chi)}=\eta\,{\hat n}_{\rm d}-\frac{1-\eta}{N}
\ {\hat Q}_{\chi}\cdot{\hat Q}_{\chi}\ .
\label{Alhassid}
\end{equation}
Here, two control parameters $\eta$ and $\chi$ change within the
bounds $0\le\eta\le 1$ and $-\sqrt{7}/2\le\chi\le 0$.
If $\eta$=1, the hamiltonian (\ref{Alhassid}) has the U(5) symmetry,
while with $\eta$=0 the hamiltonian has the SU(3) symmetry for
$\chi$=$-\sqrt{7}/2$ or the SO(6) symmetry for $\chi$=0.
For other parameter values, the hamiltonian has no dynamical
symmetry, i.e., is transitional between two or more limits.

The statements concluding the last paragraph can be particularly easily
read in the following expansion of Eq.\,(\ref{Alhassid}) into Casimir
operators from Tab.\,I:
\begin{eqnarray}
{\hat H}_{(N,\eta,\chi)} & = & \left [ \eta-\frac{1-\eta}{N}\left(\frac{\chi}
{\sqrt{7}}+\frac{2\chi^2}{7}\right)\right ]\,{\hat{\cal C}}_1[{\rm U(5)}]-
\frac{1-\eta}{N}\left(\frac{\chi}{\sqrt{7}}+\frac{2\chi^2}{7}\right)\,
{\hat{\cal C}}_2[{\rm U(5)}]-
\nonumber\\ & &
-\frac{1-\eta}{N}\left(1+\frac{2\chi}{\sqrt{7}}\right)\,
{\hat{\cal C}}_2[{\rm SO(6)}]+
\frac{1-\eta}{N}\left(1+\frac{3\chi}{\sqrt{7}}+\frac{2\chi^2}{7}\right)\,
{\hat{\cal C}}_2[{\rm SO(5)}]+
\nonumber\\ & &
+\frac{1-\eta}{N}\frac{\chi}{\sqrt{7}}\,{\hat{\cal C}}_2[{\rm SU(3)}]-
\frac{1-\eta}{N}\left(\frac{\chi}{\sqrt{7}}+\frac{\chi^2}{14}\right)\,
{\hat{\cal C}}_2[{\rm SO(3)}]\ .
\label{expansion}
\end{eqnarray}
What we further see from this expansion is that while the hamiltonians
with $\chi$=0 or $\chi$=$-\sqrt{7}/2$ ($\eta$ varying) are purely
U(5)--SO(6) or U(5)--SU(3) transitional, respectively (they contain no
admixture of the SU(3) or SO(6) Casimir operators, respectively), the
$\eta$=0 ($\chi$ varying) case represents the SU(3)--SO(6) transition
with some amount of the U(5) \lq\lq impurity\rq\rq\ (see a schematic
illustration in Fig.\,1).

The hamiltonian (\ref{Alhassid})=(\ref{expansion}) reaches neither the
SU(3)* nor the SO(6)* symmetry for any values $(\eta,\chi)$
\cite{Cejnar}.
However, one could easily write another two-dimensional parameterization
that would pass the SU(3)* and/or SO(6)* dynamical symmetries somewhere
in the middle of the Casten triangle (more precisely, the SU(3)* can
indeed lie in the middle, but the SO(6)* can only be on the U(5)--SO(6)
edge \cite{Kusnezov,Cejnar}).
One therefore has to be very careful in generalizing results obtained
within a single parameterization.
If, as an example, the SU(3)* and/or SO(6)* dynamical symmetries are,
in the special parameterization, reached for some intermediate values
of the control parameters, the hamiltonian would unexpectedly receive
quite regular properties in a seemingly transitional region.
This behaviour, however, would never be encountered in an ordinary
parameterization.

We conclude this section with a few remarks on the classical analogue
of the quantal hamiltonian (\ref{Alhassid}).
As argued in Refs.\cite{Alhassid3,Whelan1}, the on the first view
incomprehensible factor $1/N$ in Eq.\,(\ref{Alhassid}) is a consequence
of the quantum--classical correspondence.
The dynamics of the IBM-1 classical counterpart is reached for $N\to\infty$
(for instance, coherent states stop overlapping in this limit) after
appropriate rescaling of the dynamical variables and hamiltonian
parameters.
It is clear that matrix elements of the quadratic Casimir invariants
(having the ${\cal O}(N^2)$ behaviour) completely dominate the ones
of linear invariants ($\sim {\cal O}(N)$) in the $N\to\infty$ limit,
which means that quantum IBM-1 hamiltonians differing only in linear
terms have all the same classical analogue.
This common ambiguity of the quantum-to-classical transition would
be very unpleasant here, because for chaos the classical behaviour is
constituent.
Fortunately, the difficulty can be bypassed by the $1/N$ damping of
the quadratic terms.
In fact, such modified hamiltonian is not a \lq\lq textbook\rq\rq\
IBM-1 hamiltonian (it would contain the uneasy operator $1/{\hat N}$
if acting in the general Fock space), but it does not matter in our
case since $N$ has always a sharp value.
In this modified model, the sets of eigenvalues and corresponding
eigenvectors for various $N$ determine quantum aspects of
classically the same system, similarly as the 3D-oscillator states with
finite quantum numbers $N$ ($=n_x+n_y+n_z$) all issue from the unique
classical origin (which the quantum system imitates in the $N\to\infty$
limit).

Following the procedure described in Ref.\cite{Alhassid3}, one derives
the classical hamiltonian $H(p_i,q_i)$ (with $i$=1\dots5) corresponding
to Eq.\,(\ref{Alhassid}).
The coordinates $q_i$ can be identified with the quadrupole-shape
variables $\beta$,$\gamma$ and Euler angles, known from the nuclear
liquid-drop model, and $p_i$ are associated momenta.
The classical potential, i.e., the function $H(p_i$=$0,q_i)$, looks as
follows (cf.\,Ref.\cite{Hatch}):
\begin{equation}
V_{(\eta,\chi)}(\beta,\gamma)=\left[\frac{5}{2}\eta-2\right]\beta^2+
\left[(1-\eta)\left(1-\frac{\chi^2}{14}\right)\right]\beta^4+
\left[\frac{2}{\sqrt{7}}(1-\eta)\chi\right]\beta^3
\sqrt{1-\frac{\beta^2}{2}}\cos(3\gamma)\ .
\label{potential}
\end{equation}
This function is shown, for some particular $(\eta,\chi)$ values, in
Fig.\,2.
In agreement with a general rule, one sees that the U(5), SU(3) and
SO(6) limits correspond to, respectively, spherical ($\beta$=0,
$\gamma$ arbitrary), deformed axially symmetric ($\beta$$\neq$0,
$\gamma$=0) and deformed \lq\lq $\gamma$-soft\rq\rq\ ($\beta$$\neq$0,
$\gamma$ arbitrary) shapes at the minimum potential energy.
A \lq\lq phase transition\rq\rq\ \cite{Iachello} from deformed
to spherical shape can be observed for $\eta$=4/5, where the potential
develops a minimum at $\beta$=0.
Although all features of the IBM-1 classical limit certainly cannot be
derived from the potential alone (kinetic terms of the Hamilton function
also have a rather specific form), the $\eta$=4/5 value can be, within some
plausible simplification, seen as a border between various classical
modes of the model (a more sophisticated approach was recently described
in Ref.\cite{Lopez}).

\section{Wave-function entropy}
\label{entropy}

\subsection{Definition}
\label{definition}

The Shannon information entropy of wave functions
\cite{Blumel,Iachello1,Izrailev,Reichl,Whelan2,Jolie,Zelevinsky1,Cejnar}
is a natural measure of the localization of wave functions in a given
basis.
The wave-function entropy (as we will briefly call it) of a state
$\ket{\psi}$ with respect to the basis ${\cal B}\equiv\{\ket{i^{\cal B}}\}
_{i=1}^n$ ($n$ is the dimension of the Hilbert space) is defined by the
following formula:
\begin{equation}
W_{\psi}^{\cal B}=-\sum _{i=1}^n|a_{\psi i}^{\cal B}|^2\ln |a_{\psi i}
^{\cal B}|^2\ ,
\qquad\qquad a_{\psi i}^{\cal B}=\langle i^{\cal B}|\psi\rangle\ .
\label{entrop}
\end{equation}
It is minimum ($W_{\psi}^{\cal B}=0$) if $\ket{\psi}$ coincides with one
of the basis vectors, while the maximum ($W_{\psi}^{\cal B}=\ln n$) is
reached if $\ket{\psi}$ is spread uniformly among all basis states, i.e.,
if $|a_{\psi i}|^2=1/n$.
The intermediate entropy values indicate a partial fragmentation of the
state $\ket{\psi}$ in the basis $\cal B$.

\subsection{Reference bases}
\label{entropybas}

An apparent question accompanying the use of the wave-function entropy
concerns an appropriate selection of the reference basis.
Of course, this selection must issue from the physical aims followed.
If effects of some perturbation on the system are to be measured, the
reference basis will naturally be the eigenbasis of the unperturbed
hamiltonian.
Recent studies of Zelevinsky {\it et al.\/}
\cite{Zelevinsky1,Zelevinsky2,Zelevinsky3,Zelevinsky4} provide an
interesting example.
In these works, the $sd$-shell model with residual two-body interactions
was considered, and \lq\lq dissolving\rq\rq\ of the actual eigenstates
in the shell-model basis was measured for a system of 12 fermions.
It was shown that for realistic strengths of residual interactions the
relative wave-function entropy of individual states along the spectrum
follows almost exactly the state-density logarithm, which allows one
to relate the wave-function and thermodynamic entropies.
For bases unrelated to the unperturbed hamiltonian, however, the entropy
was equally large for all states (just like in the case of too strong
residual interactions) indicating the irrelevance of the basis selected.

In our case, selection of the reference bases naturally follows from
the demand to measure a \lq\lq similarity\rq\rq\ of a given
general hamiltonian to the dynamical-symmetry limits.
Associated with each integrable (dynamically symmetric) system is
a set of mutually commuting integrals of motion, which (in a favourable
case) defines a single physically important basis or (in general)
a subclass of bases.
The basis is not always unique because the system can allow for
building more independent sets of integrals of motion.
The IBM-1 is a good example:
In this case, the dynamical-symmetries contain some missing labels
\cite{Frank,Iachello}, which means that Casimir operators of any
given chain must be supplied by some other operators to form 
a complete set of commuting integrals of motion \cite{Zhang2,Zhang4}.
For various choices of the additional commuting operators one obtains
generally different bases.
All these are eigenbases of the corresponding dynamical-symmetry
hamiltonian, and the ambiguity of the basis selection can be seen
as resulting from unavoidable spectral degeneracies in the IBM-1
limits due to the missing labels.
Nevertheless, we will see later that in the cases studied these
ambiguities are not large enough to paralyze predictions based on
a particular choice of the dynamical-symmetry basis.

It must be emphasized that the constraints on the basis in the
Hilbert space are the only signatures of a particular dynamical
symmetry (or a particular case of integrability).
Indeed, as can be easily evidenced, the spectrum corresponding
to an integrable system can be made arbitrary without changing
integrals of motion (i.e., preserving the integrability) if
only the eigenbasis is conserved.
Also any given dynamical symmetry can be associated with an
arbitrary spectrum since the symmetry itself does not impose any
constraints upon the way in which the hamiltonian depends on
the associated Casimir invariants.
A U(6)-generated boson hamiltonian, for instance, with a given
dynamical symmetry does not have to follow the special form
assumed in the IBM-1, but could be an arbitrary function of
general-order Casimir operators of the given group chain.
The non-generic spectral properties of various integrable systems,
such as the non-Poissonian level spacing distribution recently
noticed \cite{Paar} and explained \cite{Lauritzen} in the IBM-1
for a particular SU(3) hamiltonian, e.g., seem to illustrate
these matters.

In this work, we use the wave-function entropy to quantify the
departure of transitional IBM-1 hamiltonians in Eq.\,(\ref{Alhassid})
from the particular dynamical symmetries of the model.
The bases $\cal B$ of interest will thus be associated with the limits
(\ref{chains}) and (\ref{chains*}) and we will deal with ${\cal B}\equiv$
U(5), SU(3), SO(6), SU(3)*, and SO(6)* entropies.
As pointed out above, the IBM-1 dynamical-symmetry bases are not
unique due to the degeneracy caused by missing labels, which is
a problem that must be solved first.
To do it precisely, one should evaluate each of the above five
entropies in all bases allowed by the respective symmetry and
discuss results with regard to the obtained uncertainty intervals.
Nevertheless, it can be directly shown that the uncertainties
should not be very large in the cases studied because of a relatively
low average multiplicity of degeneracies.
In fact, one has to consider what amount of the wave-function
amplitudes in a given dynamical-symmetry basis remains uncertain
due to the basis ambiguity.
If 100\,\% is assigned to a completely degenerated hamiltonian and
0 to a quite non-degenerated one, the following quantity can be used
to measure this amount:
\begin{equation}
f_{\rm deg}=\frac{\sum_i n_{\rm deg}^{(i)}(n_{\rm deg}^{(i)}-1)}
{n(n-1)}\ .
\label{ambig}
\end{equation}
Here, $n_{\rm deg}^{(i)}$ is the dimension of the $i$-th degenerated
subspace and $n$ is the total dimension.
The values of $f_{\rm deg}$ for various angular-momentum ($L$) subspaces
and various boson numbers ($N$) are shown in Tab.\,II.
Note that we have in mind here only the inherent degeneracy due to
missing labels, not accidental degeneracies resulting from a special
form of the hamiltonian, such as the degeneracy caused by the absence
of the SO(5) and SO(3) terms in the U(5) limit of Eq.\,(\ref{Alhassid}).
It is evident from Tab.\,II that even in the SU(3) and SU(3)* limits,
which are most degenerated, the uncertainty is still relatively small.
We therefore do not proceed in the above-proposed accurate way, but
select a single basis for each limit.
Namely, for evaluating the U(5) entropy we used the standard basis
$\{\ket{[N]n_{\rm d}vn_{\Delta}L}\}$ \cite{Frank,Iachello}, while
bases for the other limits were determined by a numerical
diagonalization of the respective hamiltonian matrices in the standard
U(5) representation (which means that the bases ${\cal B}$$\neq$U(5)
were not the standard ones).

\subsection{Average entropy}
\label{averagent}

To measure the dynamical-symmetry content of a given transitional
IBM-1 hamiltonian ${\hat H}_{(N,\eta,\chi)}$, we will average the
wave-function entropy defined in Eq.\,(\ref{entrop}) over all
eigenstates $\{\ket{\alpha,L}_{(N,\eta,\chi)}\}_{\alpha=1}^{n(L,N)}$
(with fixed angular momentum $L$) of ${\hat H}_{(N,\eta,\chi)}$:
\begin{equation}
W^{\cal B}(L,N,\eta,\chi)\equiv W^{\cal B}=\frac{1}{n(L,N)}
\sum_{\alpha=1}^{n(L,N)}W_{\alpha}^{\cal B}(L,N,\eta,\chi)
\label{average}
\end{equation}
Here, $W_{\alpha}^{\cal B}(L,N,\eta,\chi)$ is the single-state
wave-function entropy of $\ket{\alpha,L}_{(N,\eta,\chi)}$.
The average entropy $W^{\cal B}$ expresses how much the whole
eigenbasis (for the fixed $L$) of the hamiltonian under study
overlaps with the given dynamical-symmetry reference basis.

Note that now the problem with degeneracies reappears.
This time it does not concern the reference hamiltonians, but the
tested hamiltonian ${\hat H}_{(N,\eta,\chi)}$ which is at some
values of $\eta$ and $\chi$ also degenerated, leaving the
corresponding average entropy uncertain.
However, in the cases studied this uncertainty must undoubtedly be
very small, and we will neglect it in the following.

More serious is another problem:
To enable one to compare average entropies of eigenvector sets with
various dimensions $n(L,N)$, it is necessary to develop a proper
normalization.
As we already pointed out, the maximum entropy of a single state in an
$n$-dimensional Hilbert space is $\ln n$.
For $n$=2, this entropy is shared by both eigenvectors of a non-diagonal
hamiltonian in case of the maximum mixing (equal diagonal elements of
the hamiltonian).
For $n>2$, however, the entropy of a set of hamiltonian eigenstates
is influenced by the orthogonality constraints.
To calculate the maximum average wave-function entropy for $n>2$
orthonormal states becomes tricky, but there is no doubt that the result
is smaller than $\ln n$ and depends on $n$.

We solve the above problem by adopting the approach of random-matrix
theory \cite{Mehta}.
The idea is to take a set of $n$ fixed orthonormal vectors in the
$n$-dimensional space and to expand them in various orthonormal bases,
created by random rotations of the initial frame.
The resulting distribution of the average entropy $W$ depends on $n$ and
its average, for instance, may provide the desired normalization
quantity.
This procedure can be easily realized with the Gaussian Orthogonal
Ensembles (GOE) \cite{Mehta}.
By randomly generating $n$-dimensional matrices with the GOE constraints,
the above distribution is obtained from values of $W$ assigned to the set
of $n$ orthonormal eigenvectors.

For large dimensions $n$, the GOE average $\langle W\rangle_{{\rm GOE}n}$
of the entropy $W$ can be expressed explicitly:
\begin{equation}
\langle W\rangle_{{\rm GOE}n}\approx-\sqrt{\frac{2n^3}{\pi}}
\int_0^1x^2\ln(x^2)\exp\left(-\frac{nx^2}{2}\right)
{\rm d}x\approx\ln({\rm 0.482\,}n)
\label{GOE}
\end{equation}
(see also Ref.\cite{Zelevinsky4}).
For small $n$, however, serious deviations from this value can be expected
due to non-Gaussian distributions of the eigenvector components.
We performed a Monte--Carlo simulation whose results are shown in Fig.3.
The ensemble average $\langle W\rangle_{{\rm GOE}n}$ and the band which
contains 90\,\% of the $W$ distribution are both shown in Fig.3a for
dimensions $n$=2---30.
As can be seen, $\langle W\rangle_{{\rm GOE}n}$ is considerably lower than
$\ln n$.
The width of the entropy distribution decreases with $n$, which is
demonstrated also by the relative r.m.s. deviation $d$ of the entropy
$W$ from the GOE average in Fig.3b.
In Fig.3b, in addition, the coefficient $\alpha _n$ from a parameterization
$\langle W\rangle_{{\rm GOE}n}=\ln(\alpha_nn)$ is displayed.
The convergence of $\alpha _n$ to its asymptotic value from Eq.(\ref{GOE})
is evident.

For any set of $n$ orthonormal vectors, the average entropy $W^{\cal B}$
in a {\em randomly chosen\/} basis $\cal B$ will most probably be close to
$\langle W\rangle_{{\rm GOE}n}$
(cf.\cite{Zelevinsky1,Zelevinsky2,Zelevinsky3,Zelevinsky4}).
It is therefore reasonable to normalize the entropy values to the GOE
average.
We emphasize, however, that the fraction $W^{\cal B}/\langle
W\rangle_{{\rm GOE}n}$ can be even larger than 1.
To be absolutely exact, one should take into account also the fact
that the range of probable deviations of the GOE-normalized entropy
from unity depends on $n$ since the shapes of the entropy
distribution for various dimensions differ.
The deviations can be characterized by the probability localized below
(or above) $W^{\cal B}$ (see, for instance, the lower and upper 5\,\%
limits displayed in Fig.3a as the \lq\lq GOE 90\,\%\rq\rq\ band).
Therefore, the GOE distributions of the wave-function entropy can,
in principle, induce a kind of metric for measuring the proximity of
a given basis to another one.

\subsection{Entropy-ratio product}
\label{entropyrat}

In the following, we will most often display the entropy ratio
\cite{Cejnar}
\begin{equation}
r^{\cal B}=\frac{\exp W^{\cal B}-1}{\exp\langle W\rangle_{{\rm GOE}n}-1}
\ .
\label{ratio}
\end{equation}
Note that the quantity $n_{{\rm eff}\psi}^{\cal B}=\exp W_{\psi}^{\cal B}$
can be understood as an effective number of wave-function components
of the state $\ket{\psi}$ in the basis $\cal B$ (it changes between 1
and $n$).
An analogous quantity, $n_{\rm eff}^{\cal B}=\exp W^{\cal B}$, derived
from the average entropy is some \lq\lq average\rq\rq\ (not the rigorous
statistical average) effective number of wave-function components assigned
to a given eigenvector set.
The fraction $\exp W^{\cal B}/\exp\langle W\rangle_{{\rm GOE}n}=
n_{\rm eff}^{\cal B}/\alpha_nn$ has a better \lq\lq contrast\rq\rq\ than
the GOE-normalized entropy $W^{\cal B}/\langle W\rangle_{{\rm GOE}n}$,
but its lower bound depends on $n$.
The quantity in Eq.\,(\ref{ratio}) has both a good contrast and a constant
(=0) lower bound.

To express simultaneously the average overlap of a given eigenbasis
with all the reference bases, we introduce the product of the five
entropy ratios from Eq.\,(\ref{ratio}):
\begin{equation}
R=C \prod_{{\cal B}={\rm U(5)\dots SO(6)*}}\ r^{\cal B}\ ,
\label{product}
\end{equation}
where $C$ is an arbitrary normalization constant.
Trivially, $R$ is zero if any of the ratios $r^{\cal B}$
vanishes, while if all $r^{\cal B}$s are large ($\approx$1), $R$ is
large, as well.
This qualifies the entropy-ratio product $R$ to decide whether the
system is close (or not) to any of the five possible dynamical
symmetries regardless of what symmetry it actually is.
However, the reasoning is not as clear if $R$ is small due to
a simultaneous partial suppression of more ratios $r^{\cal B}$ in
Eq.\,(\ref{product}).
Then a question raises, whether the partial influence of several
individual symmetries on the system's behaviour is really
\lq\lq cumulative\rq\rq\ in such a case (as implicitly assumed
in the construction of $R$) or whether one should not take into
account only the nearest symmetry (i.e., to use the minimum of
the five ratios $r^{\cal B}$ instead of $R$).
Based on a good correlation of the product $R$ with standard
chaotic measures, as it will be presented below, we incline to
believe that its definition is justified.

\section{Results}
\label{results}

In their remarkable series of works
\cite{Alhassid1,Alhassid2,Alhassid3,Alhassid4,Whelan1},
Alhassid, Whelan, and Novoselsky mapped the classical and quantum
signatures of chaos associated with the hamiltonian (\ref{Alhassid})
in the whole ($\eta$,$\chi$)-parameter range for various angular
momenta.
As shown in a recent work \cite{Cejnar}, the observed behaviour
of standard chaotic measures has a counterpart in the behaviour of
the entropy-ratio product $R$ from Eq.\,(\ref{product}).
We review these results in more detail (paragraphs C and D below)
and discuss (par.\,E) also the role of the boson number $N$.
Before (par.\,A, B) we concentrate on properties of the wave-function
entropy of individual states in some less complex cases.

\subsection{U(5)-symmetry breaking}
\label{u5}

If the value of $\eta$ goes down from 1, the hamiltonian (\ref{Alhassid})
looses the U(5) dynamical symmetry.
As shown in Sect.\,\ref{model}, the $\beta$=0 minimum of the potential
(\ref{potential}), characteristic for the U(5) limit ($\eta$=1), keeps
existing to the limiting value $\eta$=4/5, where a \lq\lq phase
transition\rq\rq\ to a deformed shape takes place in the classical
system.
It is therefore interesting to look at what happens around this critical
value with the quantum system.
In the quantum case, the U(5)-symmetry breaking can be monitored by
the wave-function entropy, which is clearly zero at the U(5) vertex,
but increases as the symmetry is being departed.
The average U(5) entropy\,---\,or, more precisely, the ratio
$r^{\rm U(5)}$ from Eq.\,(\ref{ratio})\,---\,of all states with angular
momentum $L$=0 is shown for the U(5) side of the Casten triangle,
namely for $0.5\le \eta\le 1$, in the left column of Fig.\,4.
The two histograms correspond to boson numbers $N$=10 and $N$=20 (in both
cases, the number $n$ of $L$=0 states is indicated).

The average spread of the actual eigenstates in the U(5) basis, as shown
in the left column of Fig.\,4, changes quite smoothly\,---\,no abrupt
transition appears neither at $\eta$=4/5 nor at any other value.
However, it is not quite so if the ground state alone is concerned.
The ground state is, of course, most sensitive to changes of the potential
minimum.
Indeed, the ground-state's U(5)-entropy ratio $r_1^{\rm U(5)}$, shown
in the middle two histograms in Fig.\,4., changes rapidly around
the critical $\eta$ value.
The squared amplitude modulus corresponding to the admixture of the
unperturbed U(5) ground state in the actual ground state is displayed
in the right-hand column of Fig.\,4.
It is evident that changes of the ground state become, in agreement
with general expectations \cite{Rowe,Iachello}, sharper as the boson
number $N$ increases.
These results might turn out interesting in connection with a recent
attempt to attribute a critical phase-transitional behaviour
(rotor--vibrator) to low-lying collective states in atomic nuclei
\cite{Casten}.

\subsection{U(5)--SU(3) transition}
\label{u5su3}

The IBM-1 enables one to study not only processes of a single-symmetry
breaking, but also transitions between two or more symmetries.
The entropy measures connected with both the dynamical-symmetry bases
in play then quantify breaking of one symmetry and the simultaneous
onset of a new symmetry.
Here we consider the transition between U(5) and SU(3) limits of the
hamiltonian in Eq.\,(\ref{Alhassid}), namely the way from $\eta$=1
to $\eta$=0 along the $\chi$=$-\sqrt{7}/2$ edge of the Casten
triangle (see Fig.\,1).
The U(5) and SU(3) wave-function entropies for individual $L$=10
states with $N$=20 are shown in Fig.\,5 (notice that the orientation
of axes is opposite in both histograms).
In accord with the previous paragraph, an average of the single-state
entropies gradually increases as the respective dynamical symmetry is
left.
However, the histograms in Fig.\,5 show in more detail the way in
which the symmetry breakdown proceeds:
The variable $\alpha=1\dots121$(=$n$) enumerates the eigenstates
$\ket{\alpha,L}_{(N,\eta,\chi)}$ consecutively with the increasing
energy.
When departing from a given symmetry, the mixing of states concerns
at first more the medium-energy states than the states on the spectral
tails.
This is still valid somewhere on the midway between the two symmetries.
Nevertheless, the respective entropies keep growing until a saturation
value, roughly equal to the GOE average, is reached for almost all states
except for a few ones on the tails.

This behaviour can be related to general trends following from
perturbation theory.
Eigenstates of a hamiltonian affected by a small perturbation are
mixtures of unperturbed eigenstates.
The amplitude of the $j$-th unperturbed state (with energy $E_j$) in
the $i$-th perturbed state (whose unperturbed energy was $E_i$) is
proportional to $1/|E_i-E_j|$ times the mixing matrix element.
The eigenstate density culminates (for models with finite state
numbers) in the middle of the spectrum, which means that in the
statistical case (\lq\lq random\rq\rq\ matrix elements of the
perturbation) the mixing should be maximal there, as well.
When the perturbation strength becomes large enough to mix states
from opposite ends of the spectrum, practically all wave functions
reach the GOE entropy value.

As already discussed in Sect.\,\ref{entropybas}, such scenario seems
to work also in the shell model
\cite{Zelevinsky1,Zelevinsky2,Zelevinsky3,Zelevinsky4}.
However, it must be stressed that it can be invalid in some cases,
especially if there exist some \lq\lq non-statistical\rq\rq\
structural effects along the spectrum, as illustrated in the
following example:
Consider two classes of the IBM-1 states, both being mixtures of
the U(5) eigenstates with various d-boson numbers.
Let $n_{\rm d}\le n_{\rm d0}$ for the first class of states and
$n_{\rm d}>n_{\rm d0}+2$ for the second class ($n_{\rm d0}$ is an
arbitrary number smaller than $N$$-$2).
The states from both classes cannot be mixed by a two-body
interaction, so that if they prevail in some part of the spectrum,
the wave-function entropy would be systematically reduced there
compared to the level-density expectation.

We should have in mind that in our case the above-discussed
correspondence between the complexity and density of eigenstates
must be imperfect since the changes of the IBM-1 hamiltonian
under study are not small perturbations.
The $\eta$-dependence of a smoothed state density
\begin{equation}
\rho(E)=\int\rho_{\rm e}(E')\,g(E-E')\,{\rm d}E'\ ,
\label{density}
\end{equation}
where $\rho_{\rm e}(E)$ is the exact state density (a chain of
delta-functions) and $g(E-E')$ is an appropriate zero-centered
gaussian ($\sigma$=0.07 energy units), is shown in the upper part of
Fig.\,6.
Apparently, the level distribution moves as a whole and changes in
shape under the U(5)$\to$SU(3) transition (cf. the corresponding
change of the potential in Fig.\,2).
Thus the behaviour of wave-function entropies in Fig.\,5 can be even
surprising.
Anyway, no wonder that, unlike the shell-model case
\cite{Zelevinsky1,Zelevinsky2,Zelevinsky3,Zelevinsky4}, the entropies
in Fig.\,5 cannot be directly related to the \lq\lq thermodynamic
entropy\rq\rq\ given by a logarithm of the state density.
This can be seen by comparing the histograms in Fig.\,5 with the
one in the lower part of Fig.\,6, where the smoothed state density
$\rho(E_{\alpha})$ is shown as a function of the state index
$\alpha$ (considering the different orientations of the plots in
Fig.\,5 and 6, imposed by the shape of the functions displayed).
Clearly, $\rho(E_{\alpha})$ does not exactly correspond to
$\exp W^{\cal B}_{\alpha}$.
Note, however, that dimensions $n$ in the present model are still
too low to make a definite conclusion in this question.

\subsection{U(5)--SU(3)--SO(6) transitions}
\label{triangle}

The average wave-function entropy in the whole ($\eta$,$\chi$)-range
of Eq.\,(\ref{Alhassid}) with $N$=20 was presented in our previous
work \cite{Cejnar} for various angular momenta.
It was shown that the regions with largest entropy-ratio product $R$
from Eq.\,(\ref{product}) coincide with the most chaotic regions
of the Casten triangle described in Refs.\,\cite{Alhassid2,Whelan1}
by standard quantal and classical chaotic measures.
On the other hand, the semiregular regions have the $R$-product
suppressed.

An example is in Fig.\,7, where the average-entropy ratios
$r^{\cal B}$ are shown for all five symmetries together with
the product $R$ for $L$=10 states with $N$=20 (cf. Fig.\,1 in
Ref.\,\cite{Cejnar}).
Note that because the non-standard symmetries SU(3)* and SO(6)*
are absent from the ($\eta$,$\chi$)-manifold in the present
parameterization, the corresponding entropies are never zero.
However, while the SU(3)* symmetry is totally irrelevant (it
would be present if the triangle is extended to $\chi$=$+\sqrt
{7}/2$), the SO(6)* entropy has a behaviour very similar to
U(5).

One sees in Fig.\,7 that the regular region at the U(5)--SO(6)
edge \cite{Alhassid2,Whelan1,Mizusaki} exhibits a quite high
simultaneous localization in the U(5), SO(6) and SO(6)* bases.
This is because {\em the same\/} SO(5) subgroup is common to all
the three chains I, III and III*, see (\ref{chains}) and
(\ref{chains*}), so that the $\chi$=0 hamiltonians cannot mix
states with various SO(5)-associated quantum numbers.
The block-diagonal form of the hamiltonian then naturally implies
the suppression of the above three entropies and also non-GOE
spectral characteristics \cite{Mizusaki} on the U(5)--SO(6)
transition.
As was already discussed, however, the system not only exhibits
a smaller degree of chaos but is fully integrable in this region,
since, having five degrees of freedom (for a fixed $N$; see
Sect.\,\ref{castentri}), it also has five independent compatible
integrals of motion (if not counting $\hat N$): ${\hat{\cal C}}
_2[{\rm SO(5)}]$, ${\hat L}^2$, ${\hat L}_z$, the integral associated
with the missing label in the SO(5) $\supset$ SO(3) embedding
(an invariant of SO(3) built from the SO(5) generators), and
the hamiltonian itself \cite{Alhassid2,Whelan1}.

Note that we reveal the integrable U(5)--SO(6) region using only
the bases associated with dynamical symmetries of the model,
because all the transitional bases are well localized in the
limiting ones.
That is also why we do not need to analyze wave-function
entropies of the rest of the Casten triangle in the whole
(continuous\,!) set of all integrable bases\,---\,the results
would be qualitatively the same as with the dynamical-symmetry
bases alone.
This argument holds true even in more general cases if the
integrable region with no dynamical-symmetry is located
on a transition between two chains (\ref{dynsym}) differing only
by the subgroup G$_i$, as discussed in Sect.\,\ref{symmetries}.
Finally, it should be remarked that the simple fact of the common
SO(5) subgroup along the U(5)--SO(6) transition and its dynamical
consequences remained overlooked for long \cite{Leviatan}, which
is perhaps also a reason for the potential utility of entropy
analyses.

The other non-chaotic (although not perfectly regular) region
found in Refs.\,\cite{Alhassid2,Whelan1} is the strip connecting
the U(5) and SU(3) vertices, but inside the triangle.
It is also associated with an increased localization in the
symmetry bases.
One sees in Fig.\,7 that a partial lowering of the $r^{\cal B}$
values in this region (for $\eta$ above, say, 0.5) is present in
the U(5), SO(6)*, and SO(6) histograms, and (for smaller $\eta$)
also in the SU(3) histogram.
The effect is well visible in the product histogram.
This behaviour clearly cannot be caused by some common subgroup,
as in the previous case, and its explanation is still missing.
Perhaps the newly introduced \cite{Whelan2,Alhassid5,Leviatan2},
so-called partial dynamical symmetries provide a possibility for
such an explanation.

In Fig.\,8 we compare the curve in the Casten triangle indicated
in Ref.\cite{Whelan1} as the bottom of the new semiregular valley
with the corresponding chain of boxes with a minimum value of the
entropy-ratio product $R$.
The former, evaluated as the curve of a minimal fraction $\sigma$
of the chaotic phase-space volume \cite{Whelan1}, is given essentially
by the linear function $\chi\approx[(\sqrt{7}-1)\eta-\sqrt{7}]/2$
(see Fig.\,13 in Ref.\cite{Whelan1}).
The overall agreement of the minimum-$R$ and minimum-$\sigma$ strips
is good, indicating that standard chaotic measures and the
entropy-ratio product express the same quality.
Some deviations of the two strips are probably caused by the
finite resolution of the grid in the $R$-plot, and by some
uncertainty induced by the standard chaotic measures themselves
(one could equally well use another measure than $\sigma$, yielding
probably a slightly different curve).
Also shown in Fig.\,8 are the $r^{\cal B}$ ratios and the product
$R$ on the indicated section of the Casten triangle (note that this
section does not correspond to a fixed value of $\chi$; see the
coordinate lines in Fig.\,1).
The passage of this section through chaotic and semiregular regions
can be easily identified.

\subsection{$L$-dependence}
\label{spin}

In Ref.\,\cite{Cejnar}, the wave-function entropy was evaluated for
angular momenta $L$=0, 10, 20, and 30.
If the case of $L$=30 is excluded, for which the dimension $n$
is already substantially reduced due to the proximity of the upper
angular-momentum limit $L_{\rm max}$=40 for $N$=20, the entropy
decreases with $L$ (within the given set of $L$'s) in the whole
range of the Casten triangle.
This is in agreement with the work of Alhassid and Whelan
\cite{Whelan1}, who observed an overall decrease of chaotic
measures with angular momentum.

In Fig.\,9, we present a detailed $L$-dependence of the average
wave-function entropy ratios for all momenta between $L$=0
and 20 for two particular points of the Casten triangle.
The first one ($\eta$=1/10, $\chi$=$-$$\sqrt{7}$/4) is located in
the most chaotic region, while the other ($\eta$=1/2, $\chi$=0) is
on the regular U(5)--SO(6) edge.
The previous result \cite{Cejnar}, based on the limited number of
$L$'s, is confirmed now for all $L\le 20$, but one should be aware
that at some value of the angular momentum the entropy ratio has
a minimum and turns growing (cf. Fig.\,1 in Ref.\,\cite{Cejnar}).
Note that within the chosen interval of angular momenta the
dimension $n$ changes between limits $n$=33 (for $L$=3) and
$n$=121 ($L$=8,\,10).

One sees (in the upper two diagrams in Fig.\,9) that the trend
to decrease is common to all the five entropies.
However, also apparent from Fig.\,9 is the staggering of all
entropy ratios, particularly strong for small angular momenta,
which gives rise to large oscillations in the $L$-dependence
of the entropy-ratio product $R$ (lower two diagrams).
This behaviour of the wave-function entropy, noticed already
in Ref.\,\cite{Jolie} for a different IBM-1 parameterization,
refers to an earlier observation made by Paar and Vorkapi\' c
\cite{Paar2} that within the IBM-1 the states with $L$=0,\,3
have larger spectral chaotic measures than those with
$L$=2,\,4. These findings are particularly interesting because
similar dependence was identified \cite{Abul} also in a large
experimental-data ensemble of nuclear levels.

\subsection{$N$-dependence}
\label{bosn}

Since the boson number $N$ is, like $L$, a conserved quantum number,
we shall study how the wave-function entropies vary with it.
In general, because various ${\cal C}_1[{\rm U(6)}]$=$N$ eigenspaces
carry various irreducible representations of the dynamical group, we
ask how properties of the system depend on the particular choice of
the model Hilbert space.
An example of such a dependence was already mentioned in Sect.\,\ref{u5}
(see Fig.\,4).

An important question is whether the semiregular region inside the
Casten triangle (see Sect.\,\ref{triangle}) survives when changing
$N$.
The entropy-ratio product $R$ of $L$=10 states on the triangle section
from Fig.\,8 (the line parallel with the SO(6)--U(5) side) is plotted
in Fig.\,10 for $N$=11, 14, 17, and 20.
All the $R$-plots have minima at about 3/4 of the $\eta$-range,
indicating the passage of the given section through the semiregular
region.
It seems therefore that the semiregular strip is not just a large-$N$
effect, although for $N$=11 the minimum is less pronounced relatively
to the chaotic side regions in Fig.\,10.

An interesting $N$-dependent effect appears in the U(5)--SO(6)
transitional region.
We discussed already that the U(5), SO(6), and SO(6)* entropies are
all suppressed in this region since a complete mixing of states is
disabled by the common SO(5) symmetry.
Consequently, any purely U(5)--SO(6) transitional hamiltonian has
a block-diagonal structure in the U(5), SO(6), and SO(6)* bases
\cite{Mizusaki}, each block corresponding to fixed SO(5) quantum
numbers $v$ and $n_{\Delta}$ \cite{Iachello}.
In Fig.\,11, we show the $N$-dependence of the average dimension
$n_{\rm block}$ of these blocks for $L$=10 (individual block
dimensions coincide with the degeneracy dimensions $n^{(i)}_{\rm deg}$
from Eq.\,(\ref{ambig})) relative to the total dimension $n$ of
the $L$=10 subspace.
The relative block dimension $n_{\rm block}/n$ naturally decreases
with $N$ as the number of blocks (number of allowed
($v$,$n_{\Delta}$)-values) increases, which yields an $\approx 1/N$
dependence for large $N$ and $L\ll N$.

It is clear that the decrease of the relative average block dimensions
on the U(5)--SO(6) edge with $N$ reduces increasingly also the
corresponding GOE-normalized entropies in the U(5), SO(6), and SO(6)*
bases.
Consequently, in the $N\to\infty$ limit the whole integrable U(5)--SO(6)
region would yield the above three normalized entropies equal to zero.
This is also illustrated in Fig.\,11.
The quantity that can be directly compared with the average block
dimension is the average effective number of wave-function components
$n_{\rm eff}^{\cal B}=\exp W^{\cal B}$ (see Sect.\,\ref{entropyrat}).
Open squares and triangles in Fig.\,11 indicate values of
$n_{\rm eff}^{\rm SO(6)*}/n$ for $L$=10 eigenstates at the U(5)
and SO(6) vertices.
Note that $n_{\rm eff}^{\rm SO(6)}$ at the U(5) vertex and
$n_{\rm eff}^{\rm U(5)}$ at the SO(6) vertex are determined by the
given SO(6)* values since the relation $W^{\rm SO(6)*}(L,N,\eta$=1$,
\chi$=0$)=W^{\rm SO(6)}(L,N,\eta$=1$,\chi$=0$)=W^{\rm U(5)}(L,N,
\eta$=0$,\chi$=0$)$ is valid for average entropies from
Eq.\,(\ref{average}).
Here, the first equality follows from the fact that the expansion
of the SO(6) and SO(6)* eigenstates in the U(5) basis differ only
by phases (each U(5) eigenstate has a sharp number of s- and d-bosons)
and the second equality from the evident rule that $\cal B$-expansion
matrix of $\cal B'$ is just a Hermitian conjugate of the
$\cal B'$-expansion matrix of $\cal B$.
As can be seen from Fig.\,11, the average U(5), SO(6) and SO(6)*
relative numbers of wave-function components at the U(5)--SO(6)
edge directly follow the decrease of the relative block dimensions.

\section{Concluding remarks}
\label{conc}

In this work we attempted to find a continuous measure of the
dynamical-symmetry content for a class of IBM-1 hamiltonians,
and relate it to the variety of transitional degrees between
regularity and chaos that the hamiltonians exhibit.
The key ingredient of our analysis was the simple observation
that any dynamical symmetry is connected with a certain subset
of bases, for which the average overlap with eigenbases of the
tested hamiltonians can be measured by the wave-function
entropy.
This provided us a tool for studying various phenomena
accompanying the process of dynamical-symmetry breaking
(Sect.\,\ref{results}).

We faced the following basic problems:
{\it (i)\/} Removal of the dimension-dependence of the average
wave-function entropy by normalization. It is essential if
the entropy values for subsets of eigenstates with different
conserved quantum numbers ($N$ and $L$ in our case) are to
be compared. It turned out that the GOE normalization
(Sect.\,\ref{average}) is quite satisfactory.
{\it (ii)\/} Uncertainty of the reference dynamical-symmetry
bases for non-canonical group reductions. This problem was
shown to be of minor importance for the IBM-1
(Sect.\,\ref{entropybas}), but can be more serious for other
models, for which the uncertainty should therefore be
rigorously taken into account.
{\it (iii)\/} Construction of the entropy-ratio product in
Eq.\,(\ref{product}). It remains to be an {\it ansatz}, but
seems to work reasonably well.
{\it (iv)\/} Necessity to consider also the \lq\lq hidden\rq\rq\
dynamical symmetries, such as SU(3)* and SO(6)* \cite{Cejnar}.
Note that these symmetries arise, in general \cite{Kusnezov},
from inner automorphisms of the dynamical group and are not
classified by the group theory.
{\it (v)\/} The fact that not all integrable hamiltonians of the
system are connected with dynamical symmetries. It means that
not all potentially relevant reference bases can be constructed
by group methods. Nevertheless, in the above-discussed case of
the U(5)--SO(6) transition
(Sects.\,\ref{symmetries},\,\ref{triangle}) the regular
dynamics was identified by means of only the dynamical-symmetry
bases.

The most important goal of this work was to establish a link
between the dynamical-symmetry content and the degree of
regularity/chaos.
It turned out (Sect.\,\ref{results}\,E,\,F) that for the simple
model under study the wave-function entropies are indeed
strongly correlated with the standard chaotic measures used, for
instance, in Ref.\cite{Whelan1}.
If the same conclusion can be repeated also for other dynamical
systems, the present approach would provide a new measure of
chaos, additional to the standard ones.
It is clear that the dynamical-symmetry content expressed by
the overlap of bases measures, in fact, to what extent the
integrals of motion attached to the reference dynamical
symmetry remain approximate integrals of motion for the tested
system.
In this connection, it would be interesting to know whether
also sets of some approximate or exact (?) integrals of motion,
not arising from any of the dynamical symmetries of the system,
exist and are important\footnote
{
For example, the integrable U(5)--SO(6) systems do not possess
any dynamical symmetry but their integrals of
motion\,---\,including the hamiltonian\,---\,can be constructed
solely from the integrals corresponding to dynamical
symmetries\,---\,see Eq.\,(\ref{expansion}).
}.
These questions should be addressed in next studies.

Finally, it should be stressed that we do not pretend to find
an analytical definition of any sort of generalized symmetry,
such as the partial dynamical symmetry \cite{Alhassid5}, for
instance.
Our numerical analysis allows one to see, whether the content
of a particular dynamical symmetry is small or large, but
cannot answer why it is so.
Nevertheless, even with the above limitations in mind we
believe that the approach presented in this work can yield
a new probe for investigating dynamical properties of finite
quantum systems.

\section*{Acknowledgements}
The work was supported by the internal grant No.\,38/97 of the
Charles University and by the Swiss National Science Foundation.
P.C. acknowledges an additional support from the University
of Fribourg.


\begin{table}
\caption{ Definitions of the IBM-1 operators in the convention used in
this work. }
\begin{tabular}{cc}
${\hat N}=s^{\dagger}s+d^{\dagger}\cdot{\tilde d}$
&
${\hat n}_{\rm d}=d^{\dagger}\cdot{\tilde d}$
\\
${\hat L}=\sqrt{10}[d^{\dagger}\times{\tilde d}]^{(1)}$
&
${\hat Q}_{\chi}=[d^{\dagger}\times s+s^{\dagger}
\times {\tilde d}]^{(2)}+\chi[d^{\dagger}\times{\tilde d}]^{(2)}$
\\
${\hat P}^{\dagger}_{\phi}=s^{\dagger}s^{\dagger}-e^{i\phi}
d^{\dagger}\cdot d^{\dagger}$
&
\\
\hline
${\hat{\cal C}}_1[{\rm U(5)}]={\hat n}_{\rm d}$
&
${\hat{\cal C}}_2[{\rm U(5)}]={\hat n}_{\rm d}({\hat n}_{\rm d}+4)$
\\
${\hat{\cal C}}_2[{\rm SO(6)}]={\hat N}({\hat N}+4)-
{\hat P}^{\dagger}_{\pi}{\hat P}_{\pi}$
&
${\hat{\cal C}}_2[{\rm SU(3)}]=2{\hat Q}_{-\frac{\sqrt{7}}{2}}
\cdot{\hat Q}_{-\frac{\sqrt{7}}{2}}+\frac{3}{4}{\hat L}
\cdot{\hat L}$
\\
${\hat{\cal C}}_2[{\rm SO(5)}]={\hat n}_{\rm d}({\hat n}_{\rm d}+3)-
(d^{\dagger}\cdot d^{\dagger})({\tilde d}\cdot{\tilde d})$
&
${\hat{\cal C}}_2[{\rm SO(3)}]={\hat L}\cdot{\hat L}$
\\
\end{tabular}
\end{table}

\begin{table}
\caption{
Degeneracy factors $f_{\rm deg}$ from Eq.\,(8) characterizing the
uncertainty of the IBM-1 eigenbases in the dynamical-symmetry limits
for various boson numbers $N$ and angular momenta $L$. Note that for
$L$=0 all the $f_{\rm deg}$ values are equal to 0. There is just
one $L$=30 state for $N$=15.
}
\begin{tabular}{ccccc}
        &                   & $L=10$  & $L=20$  & $L=30$   \\
\hline
 $N=15$ & U(5),SO(6),SO(6)* & 1.3\,\% & 3.3\,\% &  ---     \\
        & SU(3),SU(3)*      & 4.3\,\% & 7.5\,\% &  ---     \\
\hline
 $N=20$ & U(5),SO(6),SO(6)* & 0.7\,\% & 2.0\,\% &  3.3\,\% \\
        & SU(3),SU(3)*      & 2.5\,\% & 3.6\,\% &  7.5\,\% \\
\hline
 $N=25$ & U(5),SO(6),SO(6)* & 0.4\,\% & 1.1\,\% &  2.0\,\% \\
        & SU(3),SU(3)*      & 1.7\,\% & 2.3\,\% &  3.6\,\% \\
\end{tabular}
\end{table}

\begin{figure}
\caption{Mapping of the ($\eta$,$\chi$)-parameter space of Eq.\,(4)
onto a triangle (left) and its relation to an ideal Casten triangle
(right). Hamiltonians with non-standard dynamical symmetries (3) are
absent from the present parameterization.}
\end{figure}

\begin{figure}
\caption{Classical potential (6) as a function of quadrupole variables
$\beta$ and $\gamma$ for six ($\eta$,$\chi$) parameter pairs (A\dots F)
from  various parts of the Casten triangle.}
\end{figure}

\begin{figure}
\caption{Quantities characterizing the GOE distribution of the average
wave-function entropy for dimensions $n\le 30$ (see text).}
\end{figure}

\begin{figure}
\caption{The U(5)-symmetry breaking in the region $0.5\le\eta\le 1$
for two boson numbers (top {\it vs\/} bottom). Left: the average
U(5)-entropy ratio from Eq.\,(11) for all $L$=0 states. Middle: the
ground-state's U(5)-entropy ratio. Right: the admixture of the U(5)
ground state in the real ground state. Points outside the Casten
triangle are filled with zeros.}
\end{figure}

\begin{figure}
\caption{The U(5) (top) and SU(3) (bottom) wave-function entropy
of individual $L$=10 states along the U(5)--SU(3) transition
($\eta$=1$\to$0, $\chi$=$-(7/4)^{1/2}$). The state index
$\alpha$ is assigned increasingly with the state energy.}
\end{figure}

\begin{figure}
\caption{A smoothed density of $L$=10 states ($N$=20) along the
U(5)--SU(3) transition as a function of energy $E$ (top) and the
state index $\alpha$ (bottom).}
\end{figure}

\begin{figure}
\caption{The average-entropy ratios from Eq.\,(11) and their product
from Eq.\,(12) for $L$=10 states ($N$=20) calculated over the whole
range of the Casten triangle. The first five histograms display
$r^{\cal B}$ for the five dynamical-symmetry bases, while the lower
right histogram represents the renormalized product $R$. Like in
Fig.\,4, points outside the triangle are filled with zeros.}
\end{figure}

\begin{figure}
\caption{The semiregular region inside the Casten triangle as deduced
from classical chaotic measures (the fraction $\sigma$ of the chaotic
phase-space volume) and from the wave-function entropies (for $L$=10
states with $N$=20).
Top: The bent curve indicates the ($\eta$,$\chi$)-localization of
the $\sigma$-valley determined in Ref.\,[17], while boxes represent
local minima (if any) of the entropy-ratio product $R$ from Eq.\,(12).
Bottom: Values of $r^{\cal B}$, see Eq.\,(11), and $R$ along the given
section (the dashed line above) of the Casten triangle.}
\end{figure}

\begin{figure}
\caption{The angular-momentum dependence ($L$=0\dots 20) of the
wave-function entropy for two points of the Casten triangle (left
{\it vs\/} right). The five $\cal B$=U(5)\dots SO(6)* average-entropy
ratios $r^{\cal B}$ are shown in the upper graphs and their
product $R$ in the corresponding lower graphs.}
\end{figure}

\begin{figure}
\caption{The entropy-ratio product $R$ along the $\chi$=0.54/(1$-\eta$)
section of the Casten triangle (see the dashed line in Fig.\,8\,--\,top)
for $L$=10 and various boson numbers $N$ (cf.\,Fig.\,8\,--\,bottom).
The normalization of all $R$-plots is the same except for the $N$=11
one, which should be multiplied by 2.}
\end{figure}

\begin{figure}
\caption{The boson-number dependence of the relative average block
dimension $n_{\rm block}/n$ of the block-diagonal (in the U(5), SO(6)
and SO(6)* bases) hamiltonians at the U(5)--SO(6) transition for
$L$=10. The average relative numbers of wave-function components
$n_{\rm eff}^{\rm SO(6)*}/n$ are given separately for pure U(5) and
SO(6) hamiltonians at five values of $N$. Total numbers $n$ of the
$L$=10 states for each $N$ are indicated inside the frame.}
\end{figure}

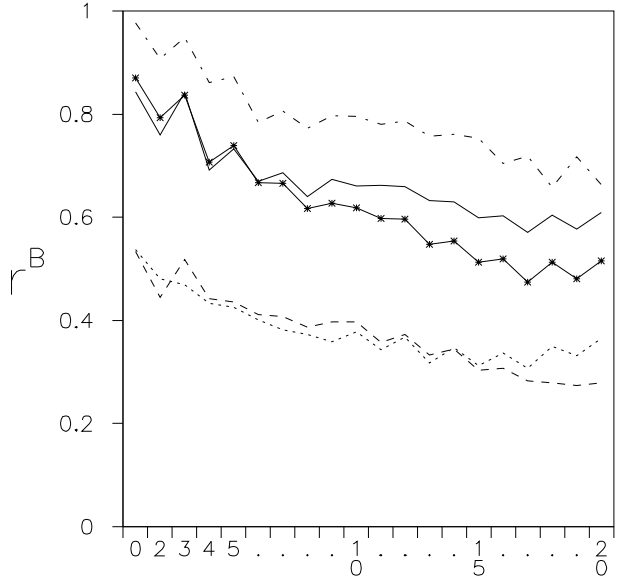
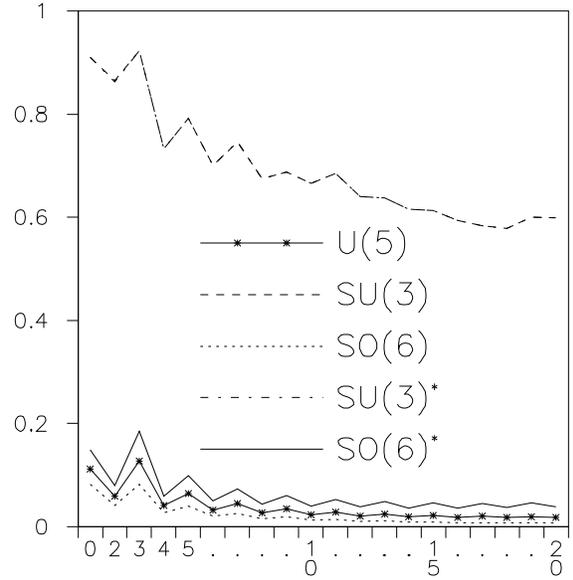
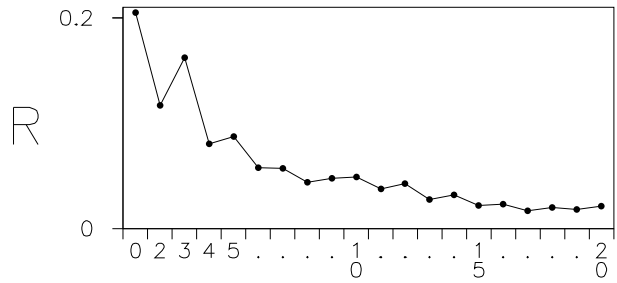
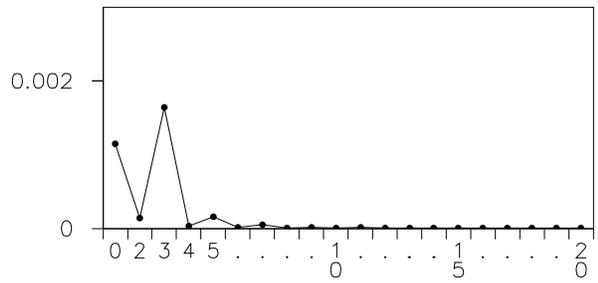
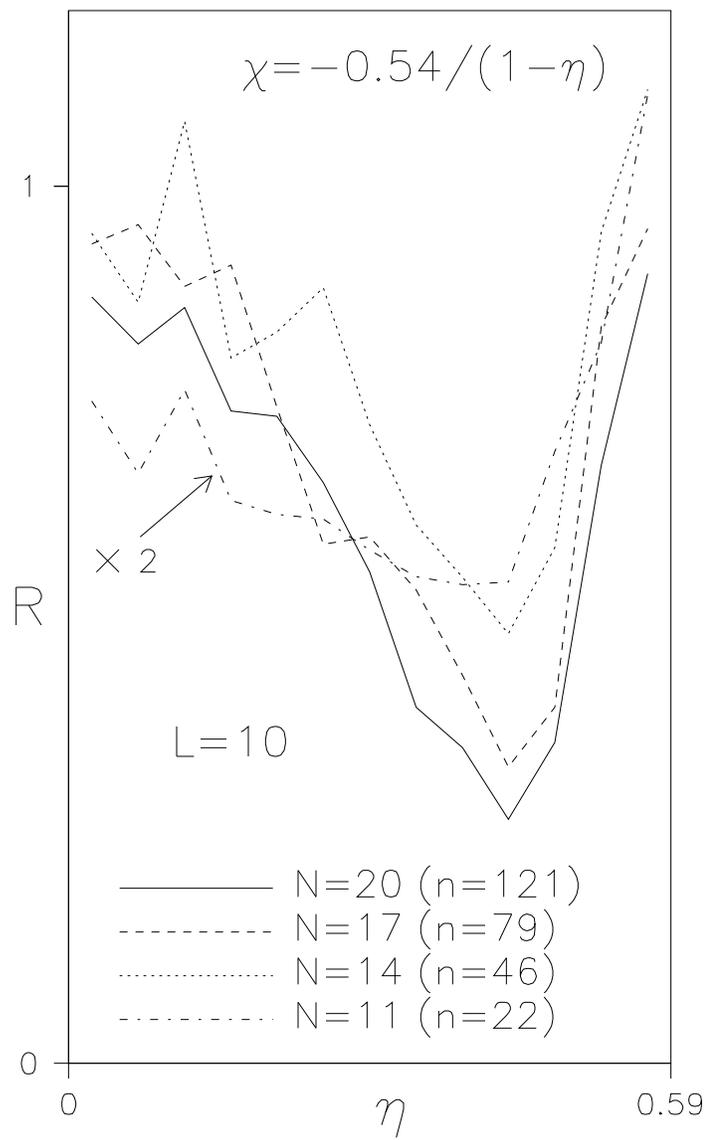
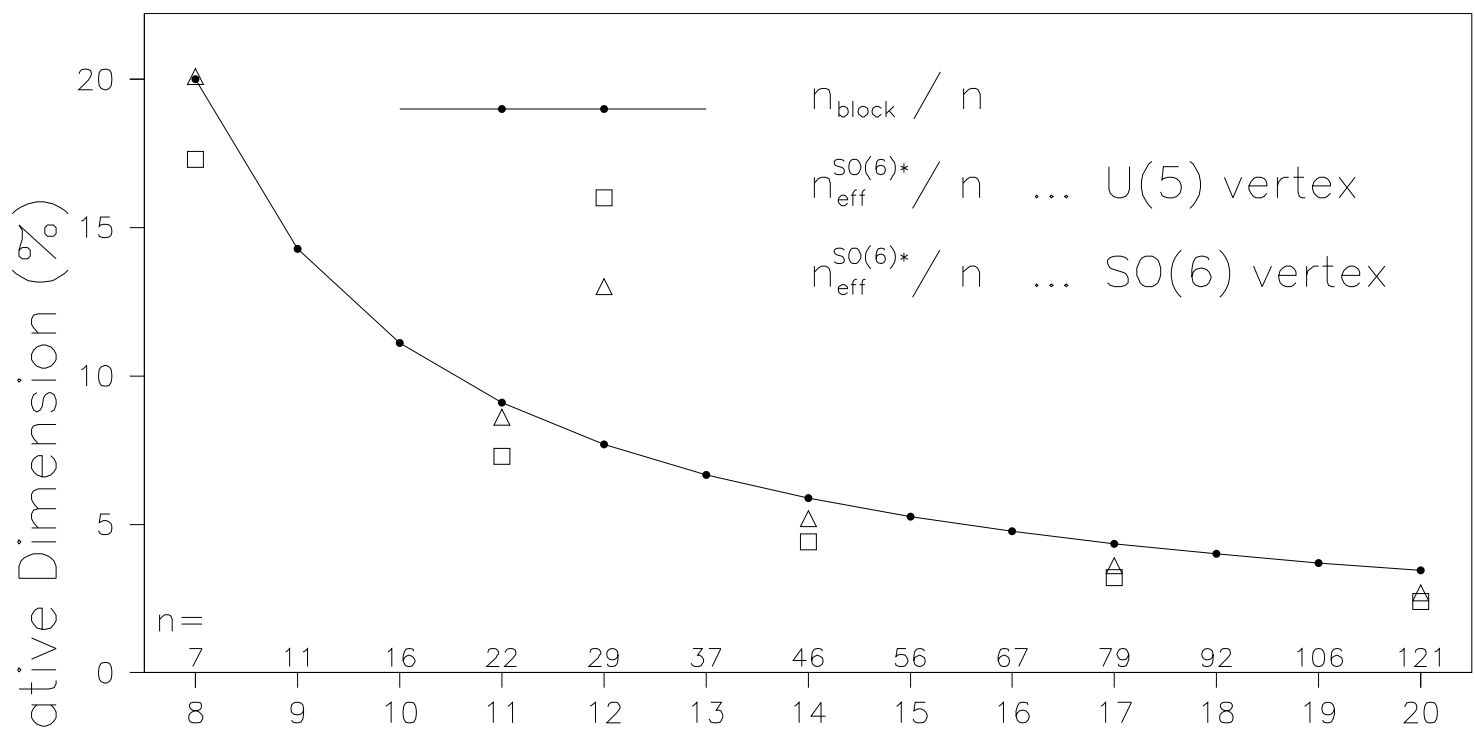
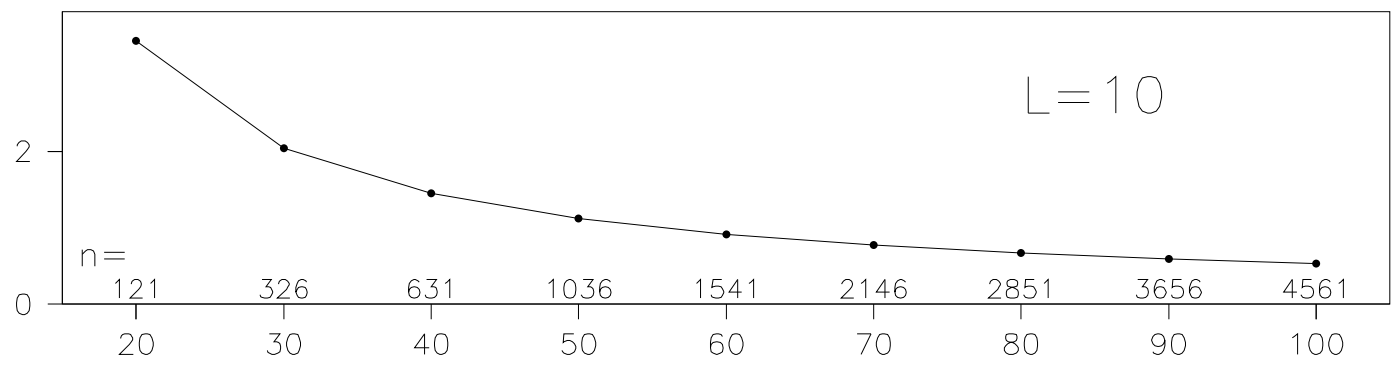
\end{document}